\documentclass[%
 reprint,
amsmath,amssymb,
 aps,
nofootinbib]{revtex4-2}

\usepackage{graphicx}
\usepackage{braket}
\usepackage{dcolumn}
\usepackage{bm}
\usepackage{xcolor}
\usepackage{slashed}
\usepackage{feynmf} 
\usepackage{cancel}
\usepackage{breakurl}
\newcommand{\comma}{,\,}
\newcommand{\semicomma}{;\,}

\begin{document}

\preprint{APS/123-QED}

\title{Symmetry, entanglement, and the $S$-matrix}

\author{Navin McGinnis}
 \email{nmcginnis@arizona.edu}
\affiliation{%
 Department of Physics, University of Arizona, Tucson, Arizona 85721, USA}%




\date{\today}

\begin{abstract}
We present a general framework connecting global symmetries to the relativistic $S$-matrix through the lens of quantum information theory. Analyzing the 2-to-2 scattering of particles of any helicity, we systematically characterize relativistic scattering amplitudes as quantum gates in the bipartite space of states with a discrete quantum number. This formalism naturally recovers and significantly extends previous results on entanglement suppression of the $S$-matrix, providing a comprehensive approach for studying the emergence of symmetries from an information-theoretic perspective. As a central result, we show that constraining the $S$-matrix to the span of minimally entangling operators is equivalent to realizing an emergent $SU(N)$ global symmetry. 
\end{abstract}

\maketitle


\noindent\textbf{\textit{Introduction}} 
- The use of symmetry as an organizing principle is a cornerstone of fundamental physics. Historically, this influence is sharply exemplified in particle physics and quantum field theory where even the notion of a \textit{particle} itself is synonymous with representations of Poincaré symmetry~\cite{Wigner:1939cj}. Yet, despite its ubiquity, it remains a long-standing question whether symmetry is fundamental or emergent from deeper physical principles.

Recent work suggests that symmetries of fundamental interactions could have an information-theoretic origin. Notably, the observed correlation of enhanced symmetries and suppressed spin-entanglement in low-energy Quantum Chromodynamics (QCD)~\cite{Beane:2018oxh}, has been formalized by interpreting the scattering of baryons as a quantum gate which minimizes the generation of entanglement~\cite{Low:2021ufv}. The consequences of minimal entanglement have since been explored further across multiple domains including low-energy QCD~\cite{Liu:2022grf,Liu:2023bnr,Hu:2024hex}, extensions of the Higgs sector of the Standard Model (SM)~\cite{Carena:2023vjc}, more general bosonic field theories~\cite{Kowalska:2024kbs,Chang:2024wrx}, and in the flavor structure of the SM~\cite{Thaler:2024anb}. The interplay between symmetries and maximal entanglement have also been explored~\cite{Cervera-Lierta:2017tdt,Miller:2023ujx}.   

These developments echo ambitions from the nascent stages of the $S$-matrix program~\cite{Cutkosky:1963oca,PhysRevLett.10.312,Sudarshan:1964cu,Leutwyler:1967wy}, which sought to derive global symmetries from first principles. Although that vision remains unrealized, it resonates with the \textit{It-from-bit} philosophy introduced much later by J. A. Wheeler (the original $S$-matrix pioneer) advocating that all aspects of Nature are information-theoretic in origin. It is compelling that this decades-old enigma from the $S$-matrix program might find resolution from information theory.

In this letter, we revisit this long-standing challenge for the $S$-matrix building on the emerging information-theoretic approach to symmetry. We develop a general, first-principles formalism to analyze the relativistic $S$-matrix as a quantum logic gate on a bipartite Hilbert space. This elegantly unifies and extends previous observations connecting entanglement and symmetry, in particular for perturbative scattering amplitudes. While previous entanglement-based studies have largely focused on symmetries consistent with free-theories, we push this perspective into the domain of more general interactions. As a central result, we will see that demanding the $S$-matrix lie within the span of minimally entangling operators implies the emergence of an $SU(N)$ global symmetry.
%

%
\noindent\textbf{\textit{$S$-matrix for qudits}} - We consider the scattering of relativistic particle states labeled by momentum, helicity and a discrete (flavor) quantum number~\footnote{The decoupling of internal quantum numbers for one particle states is guaranteed in theories with a mass gap, as spelled out in the Coleman-Mandula theorem~\cite{Coleman:1967ad}. It is widely accepted from other arguments that this is true in general~\cite{Witten:1983mna}, and we will assume this implicitly.}
\begin{equation}
\ket{p,\lambda;i}=\ket{p,\lambda}\otimes\ket{i},
\end{equation}
where the momentum-helicty states are defined in the convention of Jacob and Wick~\cite{Jacob:1959at}.
The flavor states, $\ket{i}_{i=1,..,N}$, span a finite-dimensional Hilbert space, $H_{N}\simeq\mathbb{C}^{N}$, forming a \textit{qudit} basis assumed to satisfy orthonormality and closure
\begin{equation}
    \braket{i|j}=\delta_{ij},\quad\sum_{i}\ket{i}\bra{i}=\mathbb{I}_{N}.
    \label{eq:def_rel}
\end{equation}
The one-particle states are then conventionally normalized
\begin{equation}
    \braket{p^{\prime},\lambda^\prime;j|p,\lambda;i}=(2\pi)^{3}2E_{p}\delta^{(3)}(\vec{p}-\vec{p}^{\prime})\delta_{\lambda\lambda^{\prime}}\delta_{ij}.
\end{equation}

The $S$-matrix is a unitary operator mapping asymptotic multi-particle incoming states to outgoing states. For scattering between two-particle states, its matrix elements take the form
\begin{flalign}\nonumber
    \bra{\{3\},\{4\}}S&\ket{\{1\},\{2\}}=(2\pi)^{4}\delta^{(4)}(p_{1}+p_{2}-p_{3}-p_{4})\\&\times\bra{i_{3}i_{4}}\bra{p_{3},\lambda_{3};p_{4},\lambda_{4}}S\ket{p_{1},\lambda_{1};p_{2},\lambda_{2}}\ket{i_{1}i_{2}},
\end{flalign}
where $\ket{\{i\}}\equiv\ket{p_{i},\lambda_{i};i_{i}}$, and we have assumed Lorentz invariance. The kinematic part of the $S$-matrix, $\mathbf{S}(s,\Omega)=\bra{p_{3},\lambda_{3};p_{4},\lambda_{4}}S\ket{p_{1},\lambda_{1};p_{2},\lambda_{2}}$, defines the matrix elements between momentum and helicity degrees of freedom in the usual way, and the total $S$-matrix elements
\begin{equation}
   S_{kl,ij}(s,\Omega) = \bra{kl}\mathbf{S}(s,\Omega)\ket{ij},
    \label{eq:S_tot}
\end{equation}
define the action of $S$ as a linear operator on the bipartite Hilbert space, $H_{N}\otimes H_{N}=\{\ket{i}\otimes\ket{j}\}$. 

To systematically characterize the structure of $\mathbf{S}(s,\Omega)$ we decompose its components in terms of a complete operator basis on $H_{N}\otimes H_{N}$. As a convenient choice we take the set $\{\mathbb{I}_{N}\otimes\mathbb{I}_{N},T^{a}\otimes\mathbb{I}_{N},\mathbb{I}_{N}\otimes T^{b},T^{a}\otimes T^{b}\}$, where $T^{a}$ are the generators of the special unitary group, $SU(N)$, which themselves can be constructed from the computational basis
 \begin{flalign}\nonumber
    T^k &= \frac{1}{\sqrt{2k(k+1})}\left(\sum_{i=1}^{k}\ket{i}\bra{i} -k\ket{k+1}\bra{k+1}\right),\\\nonumber
    &T^{+}(i;j) =\frac{1}{2}\left(\ket{i}\bra{j}+\ket{j}\bra{i}\right),\\
    &T^{-}(i;j) =\frac{i}{2}\left(\ket{i}\bra{j}-\ket{j}\bra{i}\right),
\end{flalign}
where $1\leq k \leq N-1$. The set of $T^{a}$ satisfy the Lie algebra brackets $[T^{a},T^{b}]=if_{abc}T^{c}$, where the normalization has been chosen so that $\text{Tr}(T^{a}T^{b})=\frac{\delta_{ab}}{2}$~\cite{Georgi:1982jb,Ramond:2010zz}. Thus, the action of the $S$-matrix on $H_{N}\otimes H_{N}$ can be parameterized as a tensor product decomposition
\begin{equation}
    \mathbf{S}(s,\Omega)=\mathcal{N}\mathbb{I}_{N}\otimes\mathbb{I}_{N}  + l_a T^{a}\otimes\mathbb{I}_{N} + r_{a}\mathbb{I}_{N}\otimes T^{a} + c_{ab}T^{a}\otimes T^{b},
    \label{eq:S_decomp}
\end{equation}
with the coefficients determined by the traces:
\begin{flalign}\nonumber
    \mathcal{N}=&\frac{1}{N^{2}}\text{Tr}(\mathbf{S}(s,\Omega)\mathbb{I}_{N}\otimes\mathbb{I}_{N}),\; l_{a}=\frac{2}{N}\text{Tr}(\mathbf{S}(s,\Omega)T^{a}\otimes\mathbb{I}_{N})\\
    r_{a}=&\frac{2}{N}\text{Tr}( \mathbf{S}(s,\Omega)\mathbb{I}_{N}\otimes T^{a}),\; \frac{c_{ab}}{4}=\text{Tr}( \mathbf{S}(s,\Omega)T^{a}\otimes T^{b}).
\end{flalign}
%

This decomposition provides an exact parametrization of the relativistic $S$-matrix as a quantum gate on the bipartite space of qudits. From the point of view of quantum information theory, a scattering process is then interpreted as a quantum operation that can be classified according to its ability to generate quantum resources, such as entanglement.

From this perspective our question becomes clear - \textit{Can symmetries of the $S$-matrix be derived or inferred when its structure as a quantum gate is constrained according to information-theoretic principles?}
In the following sections, we will show that this is indeed the case. We take an operational definition of symmetries as traceless Hermitian operators satisfying: (1) map one particle states to one particle states, (2) commute with $\mathbf{S}$, and (3) act additively on multi-particle states~\cite{Coleman:1967ad}. This set of operators can be parameterized as $Q=Q_{A}\otimes\mathbb{I}_{N} + \mathbb{I}_{N}\otimes Q_{B}$ with $Q_{A,B}$ traceless and Hermitian. 

\noindent\textbf{\textit{Minimal entanglement and conserved charges}} - To begin, we consider scattering operators which suppress the generation of entanglement when acting on the set of separable states. These operators have been fully classified for qubits ($N=2$)~\cite{Low:2021ufv}, building on foundational work on the classification of non-local qubit gates~\cite{Khaneja:2000stb,Kraus:2001ibd,Zhang:2003zz}. Additionally, the classification of operators which preserve the set of separable states for qudit systems of arbitrary (and also unequal) dimensions has been known for some time~\cite{10.1063/1.3399808,Johnston01102011,10.1063/1.3578015}.

Remarkably, operators which preserve the set of separable states can be delineated into two distinct equivalence classes associated to the identity gate and SWAP gate. Explicitly, if $\mathbf{S}(s,\Omega)$ preserves separability then it must be of the form
\begin{flalign}
    \mathbf{S}_{\sim\mathbb{I}}=(V_{A}\otimes V_{B})&\cdot \mathbf{S}_{\mathbb{I}}\cdot (W_{A}\otimes W_{B}),\label{eq:EC_id}\\\nonumber
    &\text{or}\\
    \quad \mathbf{S}_{\sim W}=(V_{A}\otimes V_{B})&\cdot \mathbf{S}_{W}\cdot (W_{A}\otimes W_{B}),\label{eq:EC_SW}
\end{flalign}
for $V_{i}$, $W_{i}\in SU(N)$, where
\begin{equation}
    \mathbf{S}_{\mathbb{I}}=\mathbb{I}_{N}\otimes\mathbb{I}_{N},
    \label{eq:id}
\end{equation}
and~\footnote{The additional factor of $2N$ compared to Refs~\cite{Low:2021ufv,Liu:2022grf,Liu:2023bnr} is due to our choice of normalization in the orthogonality condition of $T^{a}$.}
\begin{equation}
    \mathbf{S}_{W} = \frac{1}{N}\left(\mathbb{I}_{N}\otimes\mathbb{I}_{N} + 2N\sum_{a=1}^{N^{2}-1}T^{a}\otimes T^{a}\right),
    \label{eq:swap}
\end{equation}
are defined by the action $\mathbf{S}_{\mathbb{I}}\ket{ij}=\ket{ij}$, and $\mathbf{S}_{W}\ket{ij}=\ket{ji}$. In the following, we will refer to $\mathbf{S}_{\mathbb{I}}$ and $\mathbf{S}_{W}$ as the canonical members of the equivalence classes.

The classes of operators generated by Eqs~\ref{eq:EC_id} and~\ref{eq:EC_SW} effectively allow for local change of bases among the \textit{in} and \textit{out} scattering states. Thus, imposing that the $S$-matrix suppress the generation of entanglement when acting on separable states can be restated by requiring that there exists some bases where the $S$-matrix takes on one of the canonical forms
\begin{flalign}\nonumber
    \bra{kl}\mathbf{S}_{\sim\mathbb{I}(W)}\ket{ij} &= \bra{kl}(V_{A}\otimes V_{B})\mathbf{S}_{\mathbb{I}(W)}(W_{A}\otimes W_{B})\ket{ij}\\
    &=\bra{k^{\prime\prime} l^{\prime\prime}}\mathbf{S}_{\mathbb{I}(W)}\ket{i^\prime j^\prime},
\end{flalign}
reflecting the fact that local operators, written as a product of single-qudit operations, do not alter the entanglement in the system, and that the $S$-matrix acts as a map between isomorphic Hilbert spaces.
%
%
%
%

Assuming the basis in which the $S$-matrix takes on its canonical form in a given equivalence class, the associated symmetry is given by the set of generators 
\begin{equation}
    Q=(n_{A})_{a}T^{a}\otimes \mathbb{I}_{N} + (n_{B})_{a}\mathbb{I}_{N}\otimes T^{a},
    \label{eq:Q_decomp}
\end{equation}
such that $[Q,\mathbf{S}]=0$, where $n_{A,B}$ are real vectors. The symmetry of the $S$-matrix represented by any other member of an equivalence class is found by changing the basis in which $Q$ is represented accordingly.

When the $S$-matrix is in the equivalence class of the identity, $\mathbf{S}_{\mathbb{I}}$, it is clear that any $n_{A}$ or $n_{B}$ are allowed independently up to a change of basis. Thus, we obtain that the symmetry allowed by the equivalence class of the identity is $SU(N)\times SU(N)$, which is not too surprising. However, we remark this result should be interpreted as the maximal allowed symmetry.

In contrast, consider the SWAP gate, Eq.~\ref{eq:swap}. Commutation with $Q$ reduces to
\begin{flalign}\nonumber
    [\mathbf{S}_{W},Q]=&\\2\sum_{a}\sum_{b}&(n_{A})_{b}[T^{a},T^{b}]\otimes T^{a} + \sum_{c}(n_{B})_{c}T^{a}\otimes [T^{a},T^{c}].
\end{flalign}
Using the Lie algebra relations for the generators we obtain
%
%
\begin{flalign}\nonumber
    [\mathbf{S}_{W},Q]= &\\
    2\sum_{a,d}&i(\sum_{b}f_{dba}(n_{A})_{b} + \sum_{c}f_{acd}(n_{B})_{c})T^{a}\otimes T^{d}.
\end{flalign}
Due to the asymmetry in $a\leftrightarrow d$, the commutator vanishes identically so long as $n_{A}= n_{B}$. We obtain that the symmetry associated to the SWAP gate is \ the diagonal subgroup $SU(N)_{D}\subset SU(N)\times SU(N)$. 

These results extend easily to scattering on a bipartite space of unequal dimensions, $H_{N}\otimes H_{M}$~\cite{10.1063/1.3399808,Johnston01102011,10.1063/1.3578015}. In this case the $S$-matrix is realized as an element of $SU(NM)$, which always contains an $SU(N)\times SU(M)$ subgroup. Thus, the symmetry realized by the identity gate is this full subgroup, whereas for the SWAP gate we find an $SU(\text{min}(M,N))_{D}$ diagonal subgroup.

\noindent\textbf{\textit{Perturbative scattering of qudits}.} In this section, we specialize the discussion to perturbative scattering amplitudes, where we expand the $S$-matrix as
\begin{equation}
    S=\mathbb{I}+i\mathcal{T},
\end{equation}
with the kinematic part of the connected components given by $\mathbf{T}(s,\Omega)=\bra{p_{3},\lambda_{3};p_{4},\lambda_{4}}\mathcal{T}\ket{p_{1},\lambda_{1};p_{2},\lambda_{2}}$. We use the analogous tensor decomposition for the scattering amplitudes defined by
\begin{equation}
    \bra{kl}\mathbf{T}(s,\Omega)\ket{ij}=i(2\pi)^{4}\delta^{(4)}(p_{1}+p_{2}-p_{3}-p_{4})\mathcal{M}_{kl,ij}.
\end{equation}
Thus, as an operator on $H_{N}\otimes H_{N}$ we have
\begin{equation}
\mathcal{M} = \tilde{\mathcal{M}}\mathbb{I}_{N}\otimes\mathbb{I}_{N}+\mathcal{A}_{a}T^{a}\otimes\mathbb{I}_{N} + \mathcal{B}_{a}\mathbb{I}_{N}\otimes T^{a} + \mathcal{C}_{ab}T^{a}\otimes T^{b},
\label{eq:M_decomp}
\end{equation}
where
\begin{flalign}\nonumber
    \tilde{\mathcal{M}}=&\frac{1}{N^{2}}\text{Tr}(\mathcal{M}\mathbb{I}_{N}\otimes\mathbb{I}_{N}),\;\mathcal{A}_{a}=\frac{2}{N}\text{Tr}(\mathcal{M}T^{a}\otimes\mathbb{I}_{N})\\\
    \mathcal{B}_{a}=&\frac{2}{N}\text{Tr}(\mathcal{M}\mathbb{I}_{N}\otimes T^{a}), \; \frac{\mathcal{C}_{ab}}{4}=\text{Tr}(\mathcal{M}T^{a}\otimes T^{b}).
\end{flalign}
With these definitions, consider the matrix elements
\begin{equation}
    \mathcal{M}_{kl,ij} = \mathcal{C}_{ab}T^{a}_{ki}\cdot T^{b}_{lj}, \quad \text{when}\;\; k\neq i, l\neq j,
\label{eq:sel_rule1}
\end{equation}
which are responsible for generating non-local correlations between the two qudit subspaces. Appealing to entanglement suppression, we may impose $\mathcal{C}_{ab}=0$ as a sufficient  condition for $\mathcal{M}$ to act as a local operator in the flavor space. In this limit, we also have that
\begin{flalign}
    \mathcal{M}_{il,ij} =& \mathcal{B}_{a}\bra{l}T^{a}\ket{j},\quad \forall\;i\text{ and}\;j\neq l,\\
    \mathcal{M}_{kj,ij} =& \mathcal{A}_{a}\bra{k}T^{a}\ket{i},\quad \forall\;j\text{ and}\;k\neq i,
\end{flalign}
from which we obtain
\begin{flalign}
    \mathcal{M}_{il,ij}&=\mathcal{M}_{kl,kj},\quad j\neq l\label{eq:sel_rule2}\\
    \mathcal{M}_{kj,ij}&=\mathcal{M}_{kl,kj},\quad k\neq i.
    \label{eq:sel_rule3}
\end{flalign}
Further, we may write the diagonal elements as
\begin{flalign}
    \mathcal{M}_{ij,ij} &= \tilde{\mathcal{M}} + \mathcal{A}_{a}\bra{i}T^{a}\ket{i} + \mathcal{B}_{a}\bra{j}T^{a}\ket{j},\\
    \mathcal{M}_{kl,kl} &= \tilde{\mathcal{M}} + \mathcal{A}_{a}\bra{k}T^{a}\ket{k} + \mathcal{B}_{a}\bra{l}T^{a}\ket{l}.
\end{flalign}
Clearly, the sum $\mathcal{M}_{ij,ij} + \mathcal{M}_{kl,kl}$ is symmetric under the exchange of indices $j\leftrightarrow l$ or $i\leftrightarrow k$, i.e.
\begin{equation}
    \mathcal{M}_{ij,ij} + \mathcal{M}_{kl,kl} = \mathcal{M}_{il,il} + \mathcal{M}_{kj,kj.} 
    \label{eq:sel_rule4}
\end{equation}
The selections rules for the scattering amplitude that follow as a consequence of applying $\mathcal{C}_{ab}=0$ to Eq.~\ref{eq:sel_rule1}, resulting in Eqs.~\ref{eq:sel_rule2},~\ref{eq:sel_rule3}, \&~\ref{eq:sel_rule4}, elegantly recover relations arrived at previously by calculating entanglement measures at leading order in scattering amplitudes~\cite{Carena:2023vjc,Chang:2024wrx}. We emphasize that here we have shown that these relations are not merely features of specific models or approximations to various entanglement measures but, in fact, are fundamental to the structure of the $S$-matrix when $\mathcal{C}_{ab}=0$. Further, our derivation reveals that these relations are exact - unaffected by loop corrections or higher-order contributions to any specific entanglement measure.

It is important to note, though, that while $\mathcal{C}_{ab} = 0$ captures the leading order calculation of Refs.~\cite{Carena:2023vjc,Chang:2024wrx}, and is sufficient to avoid entanglement generation, it is not necessary. In Ref.~\cite{Carena:2023vjc}, it was argued that perturbative scattering amplitudes which suppress entanglement should be constrained to be in the equivalence class of the identity. However, we see that more general minimally entangling amplitudes are allowed within this equivalence class. To derive this we consider a scattering amplitude of the form
\begin{equation}
    \mathcal{M}_{kl,ij}=\bra{kl}\mathcal{M}\ket{ij}=\left[U_{A}\otimes U_{B}\right]_{kl,ij},
\end{equation} 
with $U_{A,B}$ unitary. Using Eq.~\ref{eq:M_decomp}, we impose that the correlation matrix is rank-1 and $\mathcal{C}_{ab}=\mathcal{A}_{a}\mathcal{B}_{b}$. After some manipulation, we obtain
\begin{equation}
    \text{Tr}(\mathcal{M}T^{a}\otimes T^{b})=\frac{\text{Tr}(\mathcal{M}T^{a}\otimes\mathbb{I}_{N})\text{Tr}(\mathcal{M}\mathbb{I}_{N}\otimes T^{b})}{\text{Tr}(\mathcal{M}\mathbb{I}_{N}\otimes\mathbb{I}_{N})},
\end{equation}
which can modify the selection rules of Eqs.~\ref{eq:sel_rule1},~\ref{eq:sel_rule2},~\ref{eq:sel_rule3}, \&~\ref{eq:sel_rule4}, the consequences of which will be explored in future work.

\noindent\textbf{\textit{Symmetric and symmetry-breaking interactions}} - 
We have seen that entanglement suppression constrains the structure of the scattering amplitude, leading to highly non-trivial selection rules. In this section, we make the interplay with symmetry precise by relating entanglement to the representation theory of $SU(N)$, allowing us to move beyond symmetries consistent with the identity gate. 

The defining relations for the qudits, Eq.~\ref{eq:def_rel}, are invariant under a global $SU(N)$ symmetry. Under this action, the flavor states behave as fundamental or anti-fundamental irreducible representations of $SU(N)$. Thus, there are in fact two ways to form the bipartite space of scattering qudits. In the case when one qudit transforms as a fundamental and the other as an anti-fundamental we have
\begin{equation}
    \ket{N}\otimes\ket{\bar{N}}=\textbf{1}\oplus\textbf{Adj},
\end{equation}
where $\textbf{1}$ and $\textbf{Adj}$ denote the singlet and adjoint subspaces invariant under the action of $SU(N)$. For scattering of two fundamentals we have
\begin{equation}
 \ket{N}\otimes\ket{N}=\textbf{S}\oplus\textbf{A},
\end{equation}
where $\textbf{S}$ and $\textbf{A}$ denote the symmetric and anti-symmetric subspaces. We can expand any bipartite state in the Hilbert space using projections onto the irreducible state spaces
\begin{equation}
    \ket{ij}=\sum_{R}P_{R}\ket{ij},
\end{equation}
where $R$ labels the projection onto a given irreducible representation~\cite{Mandula:1981xt}. For $(N,\bar{N})$ scattering the corresponding projections onto the singlet and adjoint subspaces are
\begin{equation}
\left(P_{\textbf{1}}\right)_{kl,ij} = \frac{1}{N}\delta_{ij}\delta_{kl}\quad \left(P_{\textbf{Adj}}\right)_{kl,ij} = \delta_{ki}\delta_{lj}-\frac{1}{N}\delta_{ij}\delta_{kl}.
\end{equation}
while for $(N,N)$ scattering, projections onto the anti-symmetric and symmetric subspaces are given by
\begin{equation}
\left(P_{\textbf{A}}\right)_{kl,ij} = \frac{\delta_{ki}\delta_{lj} - \delta_{li}\delta_{kj}}{2}\quad \left(P_{\textbf{S}}\right)_{kl,ij} =  \frac{\delta_{ki}\delta_{lj} + \delta_{li}\delta_{kj}}{2}.
\end{equation}
\begin{figure}
    \centering
    \begin{fmffile}{blob_diagram}
  \begin{fmfgraph*}(120,60)
    \fmfleft{i1,i2}   
    \fmfright{o1,o2}  

    \fmfv{decor.shape=circle, decor.filled=shaded, decor.size=50pt}{v} 

    \fmf{plain}{i1,v}
    \fmf{plain}{i2,v}
    \fmf{plain}{v,o1}
    \fmf{plain}{v,o2}

    \fmfv{label=$\ket{p_{2}\comma\lambda_{2}\semicomma\bar{\mathbf{N}}}$, label.dist=5pt}{i1}
    \fmfv{label=$\ket{p_{1}\comma\lambda_{1}\semicomma\mathbf{N}}$, label.dist=5pt}{i2}
    \fmfv{label=$\ket{p_{4}\comma\lambda_{4}\semicomma\bar{\mathbf{N}}}$, label.dist=5pt}{o1}
    \fmfv{label=$\ket{p_{3}\comma\lambda_{3}\semicomma\mathbf{N}}$, label.dist=5pt}{o2}
  \end{fmfgraph*}
\end{fmffile}\vspace{0.25cm}
    \caption{Scattering of qudits, for a pair $\ket{N,\bar{N}}$ in the initial state.}
    \label{fig:qudit_scattering}
\end{figure}
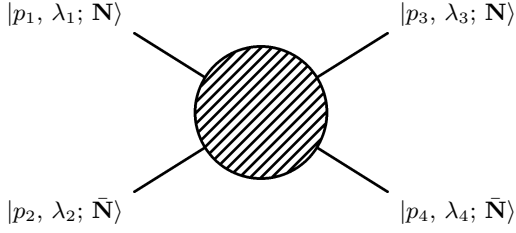
Similarly, we can expand any matrix element as 
\begin{flalign}
    \bra{kl}\mathcal{M}\ket{ij} = \sum_{R,R^\prime}\bra{mn}\mathcal{M}\ket{sr}\left(P_{R}\right)_{kl,mn}\left(P_{R^\prime}\right)_{sr,ij}.
\end{flalign}
The kinematic part of the amplitude can be systematically calculated in terms of partial waves, e.g. for $s$-channel scattering
\begin{equation}
    \bra{mn}\mathcal{M}^{(s)}\ket{sr} = \sum_{J}(2J+1)[a^{\{h\}}(s)_{J}]_{mn,sr}\mathcal{D}^{J\;*}_{\{h\}}(\Omega),
    \label{eq:partial_wave}
\end{equation}
where $\{h\}$ collectively denotes the dependence on the external helicities, and the $\mathcal{D}^{J\;*}_{\{h\}}(\Omega)$ are the Wigner D-functions~\cite{Chung:1971ri}.

We can now combine the two formalisms to systematically study the interplay between symmetry and non-local operations of the $S$-matrix. It will be useful to separate components of the amplitude which preserve an $SU(N)$ symmetry and those that break it
\begin{equation}
    \mathcal{M}=\mathcal{M}_{SU(N)} + \mathcal{M}_{\cancel{SU(N)}}.
\end{equation}
Terms in the scattering amplitude which respect the $SU(N)$ symmetry will be block diagonal in the space of irreps
\begin{equation}
    \bra{kl}\mathcal{M}_{SU(N)}\ket{ij}= \sum_{R}\mathcal{M}_{R}\left(P_{R}\right)_{kl,ij},
    \label{eq:proj_R}
\end{equation}
where we have defined 
\begin{flalign}
    \mathcal{M}_{R}&=\left(P_{R}\right)_{mn,sr}\bra{mn}\mathcal{M}_{SU(N)}\ket{sr}
\end{flalign}
For $(N,\bar{N})$ scattering in the $s$- or $u$-channel we obtain
\begin{equation}
[\mathcal{M}_{SU(N)}]^{(s,u)}_{kl,ij} = \mathcal{M}_{\textbf{1}}\left(P_{\textbf{1}}\right)_{kl,ij} + \mathcal{M}_{\textbf{Adj}}\left(P_{\textbf{Adj}}\right)_{kl,ij} ,
 \label{eq:Smatrix}
\end{equation}
while in the $t$-channel we have 
\begin{equation}
[\mathcal{M}_{SU(N)}]^{(t)}_{kl,ij} = \mathcal{M}_{\textbf{S}}\left(P_{\textbf{S}}\right)_{kl,ij} + \mathcal{M}_{\textbf{A}}\left(P_{\textbf{A}}\right)_{kl,ij},
 \label{eq:Smatrix}
\end{equation}
where our conventions for the scattering channels is provided in Fig.~\ref{fig:qudit_scattering}. For reviews on the use of projection operators for scattering amplitudes see~\cite{Mandula:1981xt,Cvitanovic:2008zz}.

In our conventions, the $s$- and $t$-channel projectors separately form an orthogonal basis for $SU(N)$-symmetric amplitudes. These bases can be related by so-called recoupling relations~\cite{Mandula:1981xt} using the Wigner 3-$j$, and 6-$j$ symbols. By choosing to reorganize the entire amplitude in the $s$-channel projector basis, we can write the total contribution of the amplitude to the correlation coefficients as
\begin{equation}
    \mathcal{C}_{ab}=\left(\mathcal{M}_{\textbf{Adj}}-\mathcal{M}_{\textbf{1}}\right)\frac{\delta_{ab}}{2}.
\end{equation}
Clearly, imposing $\mathcal{C}_{ab}=0$ constrains the amplitude to be $\mathcal{M}\sim\mathbb{I}_{N}\otimes\mathbb{I}_{N}$.~\footnote{In fact, we find $\mathcal{A}_{a}=\mathcal{B}_{a}=0$ for $SU(N)$-symmetric amplitudes implying that this condition is exact.} When symmetry-breaking terms are present, their entanglement generation can be extracted through the more general relation
\begin{equation}
    \mathcal{C}_{ab}=\text{Tr}(P_{R}\mathcal{M}P_{R^{\prime}}T^{a}\otimes T^{b}).
\end{equation}

\noindent\textbf{\textit{Emergence of $SU(N)$ symmetry}} -  In this section, we utilize our formalism to show that the selection rules corresponding to an $SU(N)$ global symmetry can be realized from an information-theoretic ansatz. To make this precise, consider an amplitude of the form
\begin{flalign}
	\mathcal{M}(s,t,u) =& \mathcal{A}(s,t,u)\mathbf{S}_{\mathbb{I}} +  \mathcal{B}(s,t,u)\mathbf{S}_{W}.
    \label{eq:span}
\end{flalign} 
From our earlier calculation it clearly follows that $[\mathcal{M},U\otimes U]=0$, where $U\otimes U$ defines the action of $SU(N)$ on the Hilbert space. Since the projection operators are themselves invariant under $SU(N)$, it follows that $[P_{R}\mathcal{M}P_{R^{\prime}},U\otimes U]=0$. The operator, $P_{R}\mathcal{M}P_{R^{\prime}}$, maps states between inequivalent invariant subspaces and commutes with the action of $SU(N)$. Thus, it follows that
\begin{flalign}
    P_{R}\mathcal{M}P_{R^{\prime}}&=0\;\text{for } R\neq R^{\prime}\\\nonumber
    &\text{and}\\
    P_{R}\mathcal{M}P_{R^{\prime}}&\propto \mathbb{I}_{N}\otimes\mathbb{I}_{N}\;\text{for } R=R^{\prime},
\end{flalign}
by Schur's lemma. The latter relation is equivalent to Eq.~\ref{eq:proj_R}.

We conclude that imposing the $S$-matrix to be in the span of the minimal entanglers, $\text{span}\{\mathbf{S}_{\mathbb{I}},\mathbf{S}_{W}\}$, provides an information-theoretic way to realize an $SU(N)$ global symmetry for the $S$-matrix.

\noindent\textbf{\textit{Conclusions}} - 
In this work, we presented a framework to study the relativistic $S$-matrix as a quantum logic gate for the scattering of particles with a discrete quantum number. By constructing a natural basis for the $S$-matrix on a bipartite Hilbert space we derived global symmetries associated to minimally entangling operators. As a byproduct, we easily recover and extend previous results on entanglement suppression and scattering amplitudes from a first-principles approach. Although our presentation has largely focused on perturbative scattering amplitudes, our results are exact and easily adaptable to more general scenarios.

Illuminating the interplay of symmetry and entanglement of scattering amplitudes leads us to an information-theoretic proposal that provides a way to resolve a decades-old issue in the relativistic $S$-matrix program~\cite{Cutkosky:1963oca,PhysRevLett.10.312,Sudarshan:1964cu,Leutwyler:1967wy}. Imposing the $S$-matrix lie in the span of minimal-entangling operators allows us to realize an $SU(N)$ interaction symmetry and derive the corresponding selection rules for scattering amplitudes.

While we have focused on entanglement correlations purely in the discrete Hilbert space, our formalism precedes a broader investigation of the interplay of entanglement and scattering amplitudes. In particular, implicit in Eq.~\ref{eq:span} is a statement on the analytic structure of the amplitude, as strictly imposing this condition at all scales of momentum transfer may enforce relations among the poles of various scattering channels, as has been seen for cases of the identity gate~\cite{Carena:2023vjc}.~\footnote{I thank Ian Low for this point.} Thus, it is of interest to investigate this aspect from the toolbox of the analytic $S$-matrix. Further, entanglement is only one aspect of quantum correlations and there is much work to be done in placing the $S$-matrix in broader context of quantum information theory~\cite{Cheung:2023hkq,Aoude:2024xpx,White:2024nuc,Low:2024mrk,Low:2024mrk,Low:2024hvn,Low:2024hvn,Liu:2025qfl,Aoude:2025ovu}.

Although we have not yet scratched the surface of a new principle in Nature, we have paved the way to explore fundamental aspects of the $S$-matrix from the perspective of quantum information.

\textit{I think, that we should admit that we do not have an understanding of the deeper causes of
any dynamic symmetry....we see only connections but not more than that.} - E. P. Wigner~\cite{Wigner1997}.
\begin{acknowledgments}
I thank Radovan Dermíšek, Andrew Jackura, Carlos Wagner and, especially, Ian Low for comments and insightful conversations. This work has been supported in part by the U.S. Department of Energy under grant No. DEFG02-13ER41976/DE-SC0009913.
\end{acknowledgments}


\nocite{*}

%

\end{document}